\documentclass[pra,twocolumn,showpacs,amsfonts,floatfix]{revtex4}
\usepackage{graphicx}
\usepackage{bm}
\newcommand{\Journal}[4]{#1 {\bf #2}, #3 (#4)}
\newcommand{\PR}{Phys. Rev.}
\newcommand{\PRL}{Phys. Rev. Lett.}
\newcommand{\PRA}{Phys. Rev. A}
\newcommand{\EPJD}{Eur. Phys. J. D}
\newcommand{\Science}{Science}
\newcommand{\JMP}{J. Math. Phys.}
\newcommand{\n}{\nonumber}
\newcommand{\bn}{\begin{eqnarray}}
\newcommand{\en}{\end{eqnarray}}
\newcommand{\h}{\hspace}
\begin{document}
\title {Theory of a one-dimensional double-X atom interferometer}
\author{Marvin D. Girardeau}
\email{girardeau@optics.arizona.edu}
\author{Kunal K. Das}
\email{kdas@optics.arizona.edu}
\author{Ewan M. Wright}
\email{Ewan.Wright@optics.arizona.edu}
 \affiliation{Optical
Sciences Center and Department of Physics, University of Arizona,
Tucson, AZ 85721}
%\author{M. D. Girardeau\thanks{Email: girardeau@optics.arizona.edu}
%and E. M. Wright\thanks{Email: Ewan.Wright@optics.arizona.edu}}
%\address{Optical Sciences Center and Department of Physics,
%University of Arizona, Tucson, AZ 85721}
\date{\today}
\begin{abstract}
The dynamics of an atom waveguide X-junction beam splitter becomes
truly 1D in a regime of low temperatures and densities and large
positive scattering lengths where the transverse mode becomes
frozen and the many-body Schr\"{o}dinger dynamics becomes exactly
soluble via a generalized Fermi-Bose mapping theorem. We analyze
the interferometric response of a double-X interferometer of this
type due to potential differences between the interferometer arms.
\end{abstract}
\maketitle
\section{Introduction}
Recent advances in atom de Broglie waveguide technology
\cite{Key,Muller,Dekker,Thy,Hinds} and its potential applicability
to atom interferometry \cite{Berman} and integrated atom optics
\cite{Dekker,Sch} create a need for accurate theoretical modelling
of such systems in the low temperature, tight waveguide regime
where transverse excitations are frozen out and the quantum
dynamics becomes essentially one-dimensional (1D) (Tonks-gas
limit). It has been shown by Olshanii \cite{Olshanii}, and also by
Petrov et al. \cite{PetShlWal00}, that at sufficiently low
temperatures and densities and high transverse frequencies
$\omega_0$, where thermal and longitudinal zero-point energies are
small compared with $\hbar\omega_0$, the transverse degrees of
freedom of a Bose-Einstein condensate (BEC) in an atom waveguide
are frozen in the ground transverse mode and the dynamics becomes
one-dimensional. If, in addition, the scattering length is large
and positive, the dynamics reduce to those of a 1D gas of hard
core, or impenetrable, point bosons \cite{Olshanii,PetShlWal00}.
This is a model for which the exact many-body energy
eigensolutions were found in 1960 using an exact mapping from the
Hilbert space of energy eigenstates of an ideal gas of spinless
fermions to that of many-body eigenstates of hard core, and
therefore strongly interacting, bosons \cite{map,map2}. Recently
two of us have extended this mapping technique in order to treat
the exact many-body dynamics of Tonks gases
\cite{soliton,breakdown} and the exact many-body ground-state of a
trapped Tonks gas \cite{1dsho}.

``Waveguide on a chip'' technology \cite{Folman,HanReiHom01} has
already advanced to achievement of BEC on a magnetic surface
microtrap \cite{Ott} and construction of a Y-beam splitter on a
chip \cite{Cassettari}, and the theory of a multimode double-Y
interferometer has been developed \cite{Andersson} in an
approximation where atomic interactions are neglected. Here we are
interested in the the opposite limit of the Tonks-gas regime
\cite{Olshanii,PetShlWal00} where interatomic interactions and
two-body correlations play a major role. Such a study is motivated
by our previous demonstration \cite{breakdown} that fermionization
strongly inhibits interference in an adiabatically split and
recombined Tonks gas. The Fermi-Bose mapping and dynamics of a
pulsed interferometer are more complicated and it is not clear a
priori whether significant interference effects will occur in this
case. This motivates the present investigation.

We set up our model for the double-X interferometer in Section
\ref{doubleX} as an effective 1D problem.
In Sec. \ref{mapping} we develop a generalized
mapping theorem and thereby map the Tonks state to a 1D Fermi gas
of free fermions, in Sec. \ref{bc}  we define the
boundary conditions at the junctions in terms of single particle
wave functions, and in the remainder of Sec. \ref{model} we
examine the limits of validity of the model and we develop a general
expression for the bosonic spatial density profiles in terms of simpler
Fermi orbitals which are not subject to the X-splitter boundary
conditions. Analytical expressions for the interferometric response
are developed in Sec. \ref{analytical} and numerical results
are presented in Sec. \ref{results}.
\section{Double-X interferometer}\label{doubleX}
The interferometer model of Andersson \textit{et al.}
\cite{Andersson} consists of an atom waveguide which is first
split into two identical waveguides by a symmetrical Y-junction,
followed by a reverse Y-junction which recombines the output of
the two arms into a single waveguide, the ``exit port'', where
interference fringes appear as a result of potential differences
between the two interferometer arms. The emphasis there was on
effects of multi-transverse-mode propagation, assuming the
transverse trap frequency low enough and/or longitudinal density
high enough and/or temperature high enough that many transverse
modes of the guides are energetically accessible and many-body
correlations are negligible. In the opposite Tonks limit where the
transverse wave function is frozen, the dynamics fermionizes and
multiple longitudinal modes play a crucial role, due to the
Fermi-Bose mapping theorem \cite{map,map2,soliton,breakdown,1dsho}
which introduces a Fermi sea of many longitudinal orbitals into
the many-boson dynamics. Rather than the double-Y geometry of
Andersson \textit{et al.}, we assume a double-X geometry since its
greater symmetry and reversibility makes implementation of unitary
straightforward; It is essentially the same model used by Scully
and Dowling \cite{SD,SZ} in their treatment of an atom waveguide
analog of a Mach-Zehnder optical interferometer. We assume that
the entrance arms, upper and lower arms after the X-splitter, and
exit arms after the X-recombiner all satisfy the necessary and
sufficient conditions for the Tonks regime
\cite{Olshanii,PetShlWal00}.
\begin{figure}
\includegraphics*[width=\columnwidth,angle=0]{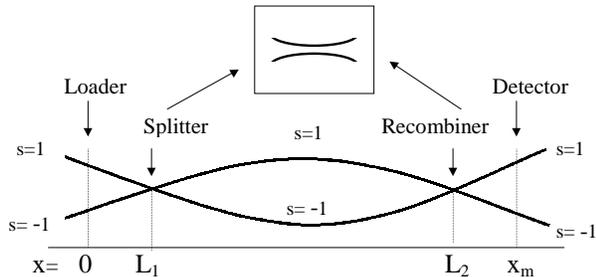}\vspace{-6cm}
\caption{Schematic diagram of the interferometer. The inset shows
that the junctions are actually avoided crossings with tunnelling
between the waveguides.} \label{Fig1.X-Tonks}
\end{figure}
We can retain an effective 1D description despite the splitting
and recombination  $(a)$ by assuming that corresponding upper
and lower interferometer
arms have the same length, so that the same longitudinal
coordinate $x$ can be used for both arms and $(b)$ by introducing
a pseudospin representation with $s=1\ (-1)$ labelling atoms in
the upper (lower) interferometer arm. Our basic model of a
double-X interferometer is shown in Fig. \ref{Fig1.X-Tonks}, and
we remark that the same model applies to an atomic loop mirror
employing an X-input-coupler. The inset indicates that the two
apparent X-crossings are really ``avoided crossings'' of the
waveguides, with tunneling between the two in their regions of
near tangency. We idealize these tunneling regions to points, see
discussion following Eq. (\ref{Tdef}), taking the X-splitter to be
at $x=L_{1}$ and X-recombiner at $x=L_{2}$ and treating the
effects of tunneling by boundary conditions specifying the wave
function discontinuities at these two points. Then $x$ specifies
the position of an atom in one of the entrance arms if $x<L_{1}$,
one of the interferometer arms if $L_{1}<x<L_{2}$, and one of the
exit arms if $x>L_{2}$. Details of these boundary conditions will
be given in the following section.
\section{Theoretical model}\label{model}
\subsection{Generalized Mapping Theorem}\label{mapping}
Our objective is to find dynamical solutions for an $N$-boson Tonks gas
in the X-splitter geometry which are exact solutions of the
time-dependent many-body Schr\"{o}dinger equation (TDMBSE) except when any
of the $N$ particle positions $x_j$ is equal to $L_1$ or $L_2$
(X-junctions), satisfy a point hard core impenetrability constraint,
have Bose symmetry (totally symmetric), and include the effect of
tunneling at the X-junctions via appropriate boundary conditions.
We do this by generalizing our previous Fermi-Bose mapping theorem
\cite{map,map2,soliton,breakdown,1dsho} to the present more complicated
geometry.

A system of N atoms is described by solutions
$\Psi(x_{1},s_{1};\cdots;x_{N},s_{N};t)$ of the TDMBSE
$\hat{H}\Psi=i\hbar\partial\Psi/\partial t$.
In the Tonks limit (large scattering length, low
linear density, tight transverse confinement)
\cite{Olshanii,PetShlWal00} the two-particle interaction behaves
as a hard core of vanishing diameter $a\to 0$. This is
conveniently treated as a constraint on allowed wave functions:
\begin{equation}\label{hardcore}
\Psi=0\ \text{if}\ s_{j}=s_{k}\ \text{and}\ x_{j}=x_{k}\ , \ 1\le
j<k\le N  ,
\end{equation}
implying that atoms in the same arm cannot interpenetrate whereas
those in different arms have no such constraint.
We start from fermionic solutions
$\Psi_{F}(x_{1},s_{1};\cdots;x_{N},s_{N};t)$ of the TDMBSE which
are antisymmetric under all space-pseudospin pair exchanges
$(x_{j},s_{j})\leftrightarrow (x_{k},s_{k})$, hence all
permutations. Generalizing the previous
definition \cite{map,map2,soliton,breakdown,1dsho}, define a ``unit
antisymmetric function" $A$ by
\begin{eqnarray}
A(x_{1},s_{1};\cdots;x_{N},s_{N})&=&\prod_{1\le j<k\le N}
\alpha(x_{j},s_{j};x_{k},s_{k}) \nonumber\\
\alpha(x_{j},s_{j};x_{k},s_{k})&=&\delta_{s_{k},s_{j}} {\rm
sgn}(x_{k}-x_{j})\nonumber \\ &+&\delta_{s_{j}1}\delta_{s_{k},-1}
-\delta_{s_{j},-1}\delta_{s_{k}1} ,
\end{eqnarray}
where $\text{sgn}(x)=+1(-1)$ if $x>0$($x<0$). For a given antisymmetric
$\Psi_F$, define a bosonic wave function $\Psi_B$ by
\begin{eqnarray}\label{bosefermi}
\Psi_{B}(x_{1},s_{1};&\cdots&;x_{N},s_{N};t)=
A(x_{1},s_{1};\cdots;x_{N},s_{N}) \n \\
&\times&\Psi_{F}(x_{1},s_{1};\cdots;x_{N},s_{N};t).
\end{eqnarray}
It satisfies the impenetrability constraint (\ref{hardcore})
because any spinless fermion wave function is space-antisymmetric
under exchange of like-pseudospin pairs. It is totally symmetric
(bosonic) under space-pseudospin permutations, obeys the same
boundary conditions as $\Psi_F$, and the Bose TDMBSE
$\hat{H}\Psi_{B}=i\hbar\partial\Psi_{B}/\partial t$ follows from
the Fermi one $\hat{H}\Psi_{F}=i\hbar\partial\Psi_{F}/\partial t$
by staightforward generalization of the previous argument
\cite{map,map2,soliton,breakdown,1dsho}, under the assumption that
the Hamiltonian is the sum of kinetic energy operators and
ordinary (non-operator) potentials. These potentials may depend on
pseudospin, i.e., they may differ between the left and right
interferometer arms, as is the case in modelling interferometric
detectors. Since the only interatomic interaction in the Tonks
limit is the point hard core treated by the constraint
(\ref{hardcore}), and this constraint is satisfied automatically
by the $N$-fermion wave functions $\Psi_F$ in (\ref{bosefermi}) as
a consequence of their antisymmetry, it follows that $\Psi_F$ is a
dynamical ideal Fermi gas state which can be written as a Slater
determinant
\begin{equation}\label{determinant}
\Psi_{F}(x_{1},s_{1};\cdots;x_{N},s_{N};t) =\frac{1}{\sqrt{N!}}\
\det_{n,j=1}^{N}\ \phi_{n}(x_{j},s_{j};t) ,
\end{equation}
where $\phi_{n}$ are $N$ orthonormal solutions of the {\em single-particle}
TDSE.
\subsection{X-splitter boundary conditions}\label{bc}
The TDMBSE is
\begin{equation}\label{TDMBSE}
\left[\sum_{j=1}^{N}\left(-\frac{\hbar^2}{2m}
\frac{\partial^2}{\partial x_{j}^2} +V_{s}(x_{j},t)\right)
-i\hbar\frac{\partial}{\partial t}\right]\Psi =0 ,
\end{equation}
where $V_{s}(x,t)$ are longitudinal potentials assumed to be
nonzero only within the interferometer arms $L_{1}<x<L_{2}$.
The mapping theorem guarantees that $\Psi_B$ will satisfy this
TDMBSE at all points $(x_{1},s_{1};\cdots;x_{N},s_{N})$ where
$\Psi_F$ does. However, $\Psi_F$ and hence $\Psi_B$ fail to
satisfy the TDMBSE at the points $x_{j}=L_1$ and $x_{j}=L_2$
where tunneling introduces discontinuities in the wave function;
at these points the TDMBSE is replaced by boundary conditions
specifying the discontinuities. Each of the orbitals in the
fermionic solution (\ref{determinant})
of the TDMBSE (\ref{TDMBSE}) satisfies the {\em single-particle} TDSE
\begin{equation}\label{TDSE}
\left[-\frac{\hbar^2}{2m} \frac{\partial^2}{\partial x^2}
+V_{s}(x,t) -i\hbar\frac{\partial}{\partial
t}\right]\phi_{n}(x,s;t)=0 ,
\end{equation}
everywhere except at the X-junctions $x=L_1$ and $x=L_2$, where this
TDSE is replaced by a boundary condition.
Since evolution according to the TDSE is unitary, the transformation
from the left to
the right side of each junction can be implemented by a unitary
$2\times 2$ matrix. Assume that at the junction half of the
probability initially in the upper arm remains in
the upper ($s=1)$ arm and half tunnels to the lower arm,
and similarly for probability entering from the lower arm,
and introduce a Pauli pseudospinor notation
\bn\label{defphi}
\bm{\Phi}_{n}(x,t)=
  \left[\begin{array}{c}
    \phi_{n}(x,+1;t)  \\
    \phi_{n}(x,-1;t)
  \end{array}\right] \quad .
\en
Then a simple boundary condition which implements the splitting at
the first X-junction is \cite{SD,Loudon}
\begin{equation}\label{boundary}
\bm{\Phi}_{n}(L_{1}+,t)=\mathbf{T}\bm{\Phi}_{n}(L_{1}-,t) ,
\end{equation}
where $L_{1}-$ ($L_{1}+$) denotes the left (right) side of the
junction and $\mathbf{T}$ is the unitary matrix
\begin{equation}\label{Tdef}
\mathbf{T}=\frac{1}{\sqrt{2}} \left(\begin{array}{cc} 1 & -1 \\ 1
& 1
\end{array}\right) \quad .
\end{equation}
Thus the amplitudes which remain in the upper or lower arms as
they pass through the avoided crossing junction suffer no phase
shifts, that which tunnels from the upper to the lower arm suffers
a phase shift of $\pi$, and that which tunnels from the lower to
the upper arm suffers no phase change. It can be shown that this
matrix can be written in the form
$e^{i\hat{V}}=(1/\sqrt{2})(1-\hat{\sigma}_{+}+\hat{\sigma}_{-})$
in terms of an interaction
$\hat{V}=i\lambda(\hat{\sigma}_{+}-\hat{\sigma}_{-})$ where
$\hat{\sigma}_{\pm}$ are Pauli pseudospin raising and lowering
operators implementing the tunneling and $\lambda$ is taken equal
to $\pi/4$ in order that probability splits equally between the
upper and lower arms. More general phase shifts could be used
subject to the proviso that $\mathbf{T}$ must be unitary, but the
simple choice here is sufficient. At the recombiner ($x=L_2$) the
same matrix boundary condition is applied once more to generate
the amplitudes in the upper and lower output arms.
\subsection{Limits of validity}\label{limits}
In our simple model the tunneling regions giving rise to the
X-splitters are replaced by point boundary conditions, and we now
examine the conditions for this to be valid. In particular, in the
boundary condition (\ref{boundary}) we treat each orbital
separately, meaning that the tunneling is assumed to proceed as
for single particles and to not be modified by many-body
interactions. For a 50:50 X-splitter, and assuming a tunneling
region length $L_{coup}$, the energy splitting $\Delta
E<<\hbar\omega_0$ between the symmetric and anti-symmetric
transverse modes of the tunnel-coupled waveguides must obey
$(\Delta E/\hbar)\cdot(L_{coup}/v_x)=\pi/4$, where $v_x$ is the
longitudinal velocity of the atoms entering the X-splitter. Here
$L_{coup}/v_x$ is the time of flight of the atoms through the
tunneling regions, and the $\pi/4$ phase-difference between the
modes is that required for a 50:50 splitter. For single-particle
tunneling to be valid we require that the energy splitting is
large compared to the many-body energy per particle of the
incoming Tonks gas of mean density $\rho$, $\Delta E>>\hbar^2\pi^2
\rho^2/6m$ \cite{map,map2}. When this condition is satisfied the
atoms tunnel individually and the transfer characteristics of the
X-splitter should therefore be the same for bosons and fermions.
For an initial gas in its ground-state which is harmonically
trapped with longitudinal frequency $\omega$ we have $\rho\approx
N/x_F$, where $x_F=\sqrt{2N}x_{osc}$,
$x_{osc}=\sqrt{\hbar/m\omega}$ being the single particle
ground-state width \cite{KolNewStr00}. In order that the tunneling
region not create large density gradients in the atomic gas we
also require $L_{coup}\ge x_F$, that is the coupling region is
longer than the width of the initial trapped gas. Furthermore,
under the assumption $L_{coup}<<(L_2-L_1)$ we may treat the tunnel
regions as points. Combining these results together, we obtain the
following conditions for the action of the X-splitters to be
treated by the boundary condition (\ref{boundary})
\begin{eqnarray}
v_x &>>& N\cdot L_{coup}\cdot\omega \\(L_2-L_1)&\gg& L_{coup}\ge
\sqrt{2N}x_{osc} .
\end{eqnarray}
We shall hereafter assume these conditions are satisfied.
\subsection{Construction of interferometer solutions}
The boundary conditions (\ref{boundary}) at the X-splitters can be
implemented as follows: Let $u_{n}(x,s;t)$ be solutions on
$x\in\left[-\infty,\infty\right]$ of the same single-particle TDSE
(\ref{TDSE}) as the $\phi_{n}(x,s;t)$ satisfy, but now \textit{no
boundary condition} is imposed on these orbitals, so they are
continuous with continuous gradient at all $x$ including $x=L_1$
and $x=L_2$. Instead, an initial condition
\begin{equation}\label{load1}
u_{n}(x,s;t=0)=u_{n}(x) ,
\end{equation}
is imposed. Note that this initial condition is independent of $s$
(the same in both entrance arms). Here $t=0$ is a time shortly
after loading of the N atoms into the entrance arm and the
$u_{n}(x)$, defined for $1\le n\le N$, are a set of orthonormal
orbitals which are negligible at $t=0$ for $x\ge L_{1}$. Then for
$t\ge 0$ and $x<L_1$, define new orbitals $\phi_n$ by
\begin{equation}\label{load2}
\phi_{n}(x,s;t)=u_{n}(x,s;t)\delta_{s1}\ ,\ x<L_1\ .
\end{equation}
These $\phi_n$, which in our model represent the actual orbitals
occupied by the atoms, satisfy the same TDSE (\ref{TDSE}) as the $u_n$
for $x<L_1$, but they are nonvanishing only in the upper arm ($s=1$),
corresponding to our assumption that the atoms are to be loaded
only into the upper arm. On the other hand, the $u_n$ are nonzero
in both the upper and lower arms as a consequence of their initial
conditions (\ref{load1}) and TDSE (\ref{TDSE}). This allows us to first
solve the TDSE without imposing boundary conditions at the junctions
$x=L_1$ and $x=L_2$, then construct solutions $\phi_n$ of the TDSE satisfying
the proper boundary conditions by defining them in terms of the $u_n$
by definitions which are properly discontinuous at the junctions.
The effect of the boundary condition at $x=L_{1}$ is such
that the upper-arm amplitude $u_{n}(L_{1},1;t)$ splits equally, with prefactors
$1/\sqrt{2}$, between the upper and lower arms at the junction.
Note that $u_{n}(L_{1},1;t)=u_{n}(L_{1},-1;t)$ since $V_s$ vanishes
for $x<L_1$ and the initial condition (\ref{load1}) is independent
of $s$. It then follows that the solutions
of the TDSE in the interferometer arms are simply
\bn\label{split}
\bm{\Phi}_{n}(x,t)=\frac{1}{\sqrt{2}}
  \left[\begin{array}{c}
    u_{n}(x,1;t)  \\
    u_{n}(x,-1;t)
  \end{array}\right] ,\quad L_{1}<x<L_2
\en
In general these $u_n$ depend on $s=\pm 1$ since the potentials $V_s$
in the arms $L_{1}<x<L_2$ do.

At the recombiner $x=L_2$ the splitting matrix $\mathbf{T}$ is
applied once more, resulting in the solutions
\bn\label{combine}
\bm{\Phi}_{n}(x,t)=\frac{1}{2}
  \left[\begin{array}{c}
    u_{n}(x,1;t)-u_{n}(x,-1;t)  \\
    u_{n}(x,1;t)+u_{n}(x,-1;t)
  \end{array}\right] , \quad x>L_2 \quad .
\en
in the output arms, where the potentials $V_s$ are again zero.

The above equations uniquely and straightforwardly determine the
many-fermion wave function $\Psi_F$ in terms of $N$ orbitals
$\phi_n$. Then within the limits of validity of our theory given
in Sec. (\ref{limits}), under which the tunneling occurs as for
single-particles and is hence the same for bosons and fermions,
the many-boson wave function is given by $\Psi_B=A\Psi_F$. Since
$A^{2}=1$, this has the important consequence that the
single-particle densities $\rho(x,s;t)$ are the same for the Bose
Tonks gas as for the ideal ``spinless'' Fermi gas of the mapping
theorem, allowing us to compute Bose dynamics and interference
fringes for this model of a Tonks gas interferometer in terms of
the corresponding solutions for the ideal Fermi gas. This will be
done in the following sections. In addition to providing the basis
for an exact calculation of the behavior of a Tonks gas of bosonic
atoms in this interferometer model via the Fermi-Bose mapping
theorem, this fermionic calculation is of interest in its own
right, since a spin-polarized Tonks gas of fermionic atoms is
dynamically equivalent to the same ideal Fermi gas inasmuch as it
satisfies the impenetrability constraint (\ref{hardcore})
automatically as a consequence of spatial antisymmetry.
\subsection{Construction of spatial density profiles}
It follows from orthonormality of the $u_{n}(x)$ and the unitarity
of Schr\"{o}dinger evolution that the $u_{n}(x,s;t)$ satisfy the
orthonormality relation
$\int_{-\infty}^{\infty}u_{n}^{*}(x,s;t)u_{m}(x,s;t)dx=\delta_{nm}$
for each fixed $s=\pm 1$. One can then show with Eqs.
(\ref{load2}-\ref{combine}) that the spinor orbitals
${\bm{\Phi}}_n$ are orthonormal in the usual sense
\begin{eqnarray} \label{orthonormality}
&&({\bm{\Phi}}_n|{\bm{\Phi}}_m)=\int_{-\infty}^{\infty}
{\bm{\Phi}}_{n}^{\dagger}(x,t){\bm{\Phi}}_{m}(x,t)dx \nonumber\\
&=&\sum_{s=\pm 1}\int_{-\infty}^{\infty}\phi_{n}^{*}(x,s;t)
\phi_{m}(x,s;t)dx \nonumber\\ &=&\frac{1}{2}\sum_{s=\pm
1}\int_{-\infty}^{\infty}u_{n}^{*}(x,s;t) u_{m}(x,s;t)dx
=\delta_{nm} ,
\end{eqnarray}
in spite of their discontinuities at the X-junctions $x=L_1$ and
$x=L_2$. Note that the orthonormality of the $u_n$ holds for each
fixed $s$ (separately in each arm), whereas the orthonormality
relation for the ${\bm{\Phi}}_n$ involves both summation over $s$
and integration over $x$. This is to be expected physically, since
the X-junctions mix probability density between both arms. Using
the Slater determinantal expression for $\Psi_F$ in terms of the
$\phi_{n}(x,s;t)$ and the fact that $|\Psi_B|^{2}=|\Psi_F|^{2}$,
it then follows that the single-particle density for both the Bose
and Fermi systems is
\begin{equation}\label{density}
\rho(x,s;t)=\sum_{n=0}^{N-1}|\phi_{n}(x,s;t)|^2 .
\end{equation}
To recapitulate, to generate the density profiles we first solve the TDSE
for each $u_n(x,s,t),n=1,\ldots,N$ in both arms, which does not
involve the X-splitter boundary conditions. Next the orbitals
$\phi_{n}(x,s;t)$ at the detector are constructed from the
boundary condition (\ref{combine}), and finally the density
profile is obtained from Eq. (\ref{density}) above.
\section{Analytical solution}\label{analytical}
H\"ansel {\it et al.} \cite{HanReiHom01} have demonstrated a
magnetic conveyor belt for atoms that could be used to launch an
atomic gas into interferometer, and recently transport of BECs by
optical tweezers has been achieved by Gustavson {\it et al.}
\cite{Gustavson}, who suggested that this technique could be used
to load a BEC into a ``waveguide on a chip''. We simulate the
tweezer loading method by choosing the $N$-fermion wave function
in the entrance or upper arm arm $s=+1$ in Fig. \ref{Fig1.X-Tonks}
as the ground-state of a harmonic trap of frequency $\omega$ with
an overall translational velocity $v_x=\hbar k/m$, $k$ being the
longitudinal wavevector. The time when the Fermi wavepacket is
released from the optical trap into the entrance arm is taken as
the time origin $t=0$ and the position of the packet-center at
that instant is taken as the space origin $x=0$. Then, the
orthonormal orbitals defining the initial condition in
(\ref{load1}) are conveniently chosen to coincide with the lowest
$N$ Hermite-Gaussians
\begin{eqnarray}\label{initialorb}
u_{n}(x)&=&C_n e^{-x^{2}/2x_{osc}^2}H_{n}\left(
\frac{x}{x_{osc}}\right )e^{ikx} \nonumber \\ C_n &=&
\frac{1}{\pi^{1/4}\sqrt{2^{n}n!x_{osc}}}  ,
\end{eqnarray}
where the N-fermion ground-state has one particle in each of the
first $N$ modes, and $x_{osc}=\sqrt{\hbar/m\omega}$ as before. The
Fermi-wavevector associated with this N-atom ground-state is
$k_F=\sqrt{2N}/x_{osc}$ \cite{RCB,DasLapWri02}, and this serves as
a measure of the wavevector on the highest occupied plane-wave
orbital. We assume throughout that $k>>k_F$ so that the
center-of-mass velocity $v_x$ of the atomic wavepacket exceeds the
expansion of the wavepacket after it is released.

Our goal in this paper is to examine the response of the double-X
interferometer to a variety of perturbations $V_s$. An interesting
choice of perturbing potentials in the interferometer arms is one
which is a delta-function pulse in time,
$V_{s}(x,t)=\hbar\Lambda_s(x)\delta(t-t_{0})$. Then, the orbitals
just after such a perturbation are related to those just before by
a spatially varying phase-shift $\Lambda_{s}(x)$:
\begin{equation}\label{t+}
u_{n}(x,s;t_{0}^{+})=e^{-i\Lambda_{s}(x)}u_{n}(x,s;t_{0}^{-})\ ,\
L_{1}<x<L_{2}\ .
\end{equation}
Such an approach was used by Rojo, Cohen, and Berman to study the
response of a Tonks gas to a delta-pulsed optical lattice
potential \cite{RCB}, and we have used it more recently to
investigate gray solitons in a Tonks gas \cite{soliton}. It is an
idealization to zero pulse length of phase imprinting techniqes
used in a number of recent experiments
\cite{Burger,Denshlag,Vogels}.

As a simple but concrete example, suppose that an external force
acting on the interferometer exerts a different impulsive force
$F_s=s\hbar[k_{p}+2Qx]\delta(t-t_{0})$ on each arm at a time $t_0$
at which the wavepackets are located within the interferometer.
(We split the impulsive force between the two arms for ease in the
analytic solution but the same physics arises if the force is
applied to one arm). The force term proportional to $k_p$
represents a spatially uniform force, whereas the term
proportional to $Q$ represents a spatially inhomogeneous force.
The corresponding potential is
$V_{s}(x,t)=\hbar\Lambda_s(x)\delta(t-t_{0})$ with
\begin{equation}\label{potential}
\Lambda_{s}(x,t)=-s[\varphi/2+k_{p}x+Qx^{2}]\delta(t-t_{0}) ,
\end{equation}
where $\varphi$ is a phase difference between the two
interferometer arms due, for example, to the Sagnac effect if the
interferometer is rotating. The factor $s=\pm 1$ means that the
impulse is oppositely directed in the left and right arms. In the
absence of a quadratic phase variation, $Q=0$, via the equivalence
principle this can be viewed as a simple model of a detector of
rotational acceleration, an ``atom gyroscope''.

\begin{figure}
\includegraphics*[width=\columnwidth,angle=0]{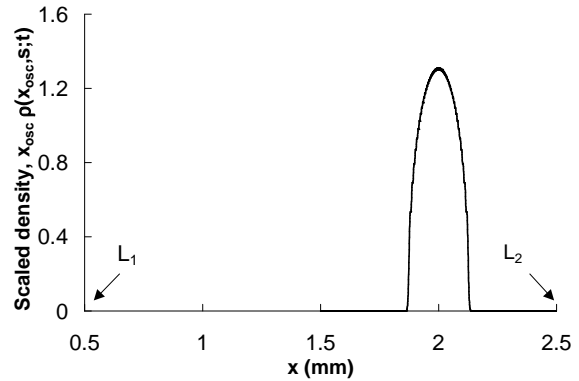}
\caption{The shape of the wavepacket in \emph{either} arm before
the pulses are applied, shown when the packet center is at
$x_{0}=2$ mm .} \label{Fig2.X-Tonks}
\end{figure}

The details of the time evolution of each orbital $u_{n}(x,s;t)$
in either arm from the time of loading until the time of detection
are given in the Appendix: The expressions for these orbitals are
somewhat cumbersome, but in general they can be written in the
form $u_{n}(x,s;t)=U_{n}(x,t)\exp(i\xi_n(x,s,t))$, where $U_n$ are
real Hermite-Gaussian functions and $\xi_n$ are phase factors, so
$|u_n|^2=|U_n|^2$. The time-evolution is the same in each arm
until the time $t=t_{0}$ when the pulses are applied. That time is
chosen so that the pulses are well contained in the interferometer
arms and the amplitude of the wavepacket at the junctions is
negligible (see Fig.~\ref{Fig2.X-Tonks}).  Thus for subsequent
propagation the extra phase (\ref{t+}) acquired by each orbital
due to the pulsed potentials can be taken to extend over all
values of $x$ although the potentials are limited to the lengths
of the interferometer arms only.

If the potentials have no quadratic spatial dependence $Q=0$, the
phase acquired is the same for each orbital, as seen from
Eq.~(\ref{linearu}) in the Appendix. In that case the densities in
the upper arm ($s=+1$) and the lower arm ($s=-1$) can be expressed
in terms of real-valued orbitals $U_{n}$ centered at
$x_{\pm}=x_{0}+\hbar(k\pm k_p)(t-t_0)/m$, see Eq. (\ref{linearu1}):
\bn\label{lindensity} \rho(x,\pm 1;t)
=\frac{1}{4}\sum_{n=0}^{N-1}\left[U_{n}^2(x-x_{+},t)+U_{n}^2(x-x_{-},t)\right]
\n\\
\mp\frac{1}{2}\cos[\theta(x,t)]\sum_{n=0}^{N-1}U_{n}(x-x_{+},t)U_{n}(x-x_{-},t)
,
\en
where $x_{0}=v_x t_0$ is the position of the center of the packets
at the scaled time of application of the pulse $\tau_{0}=\omega
t_0$, and $x_{\pm}$ the centers of the wavepackets in each arm at
the time of detection $\tau=\omega t$. The spatial
phase-modulation $\theta$ due to interference and the period
$\lambda$ of that modulation are given by
\bn\label{period}
\theta(x,t)=\varphi+2k_{p}\left[(x-x_{m})\frac{1+\tau_{0}\tau}
{1+\tau^{2}}+x_{0}\right] \n\\
\lambda=\left(\frac{\pi}{k_{p}}\right)\frac{1+\tau^{2}}{1+\tau_{0}\tau},\h{2cm}\en
where $x_{m}=(x_{+}+x_{-})/2$ is the mean position of the
recombined wavepackets at the time of detection. The exact
analytic solution (\ref{lindensity}) will be used in the following
sections to evaluate the response of the double-X interferometer
to the input atom wavepacket under varying applied perturbations.

Equation (\ref{lindensity}) for the density profile in each arm
has an illustrative limiting case. For small applied wavevectors
$k_{p}\ll k$, the centers of the wavepackets do not separate much
$x_\pm\approx v_x t$, and we have $U_{n}(x-x_{+},t)\simeq
U_{n}(x-x_{-},t)$. In this limit the density in each arm becomes
\bn\label{lindensity1} \rho(x,\pm 1;t)\simeq\frac{1}{2}\left[1\mp
\cos[\theta(x,t)]\right] \sum_{n=0}^{N-1}U_{n}^{2}(x-x_{+},t) ,\en
this expression being exact for $k_p=0$. The factor $\sum
U_{n}^{2}(x-x_{+},t)$ is the density profile of the expanding Tonks
gas, and is identical in the present approximation to that in the
absence of the applied perturbation. Thus, for $k_p\ll k$ the
response of the interferometer to the perturbation is in the
spatially dependent phase factor $\theta(x,t)$. For the case
$k_p=0$ we have $\theta(x,t)=\varphi$, so the densities in each
arm do not show spatial fringes, and we may integrate the density
over $x$ to find \emph{exact} expressions for the number of atoms
$N_\pm$ in each output arm
\begin{equation}\label{Npm}
N_\pm =\frac{N}{2}\left[1\mp \cos(\varphi)\right]  .
\end{equation}
Then if we set $\varphi=\varphi_b+\Delta\varphi$, where
$\varphi_b$ is a bias phase-shift, created, for example, using an
optical dipole potential applied to one arm, and $\Delta\varphi$
is a phase difference between the two arms due to the Sagnac
effect, we find $(N_+-N_-)=-N\cos(\varphi_b+\Delta\varphi)$.
Setting $\varphi_b=\pi/2$ we obtain
\begin{equation}\label{Deln}
(N_+-N_-)=N\sin(\Delta\varphi) .
\end{equation}
Thus, measuring the difference in numbers of atoms exiting the two
interferometer arms provides a measure of the excess phase-shift
$\Delta\varphi=2m\omega_R A/\hbar$ due to rotation, for example,
where $\omega_R$ is the rotation frequency and $A$ the enclosed
area of the interferometer.

When $k_p\ne 0$ Eq. (\ref{lindensity}) shows that the output
density in each arm can exhibit spatial fringes with period
$\lambda$ as the phase $\theta(x,t)$ becomes $x$-dependent. The
fringes become apparent when
$2x_F/\lambda=2\sqrt{2N}x_{osc}/\lambda>1$, so that there is more
than one fringe under the atomic wavepacket width. The condition
on the perturbing wavevector to observe fringes is therefore
\begin{equation}\label{con1}
\frac{k_p}{k}>\frac{1}{2\sqrt{2N}}\frac{\pi}{(kx_{osc})}\left (
\frac{1+\tau^{2}}{1+\tau_{0}\tau} \right ) .
\end{equation}
However, for momentum kicks $k_p$ such that the distance
$(x_+-x_-)=2\hbar k_p(t-t_0)/m$ between the wavepacket centers in
the two arms becomes greater than the width $x_F\sqrt{1+\tau^2}$
of the expanding Tonks gas at scaled detection time $\tau=\omega
t$, then the fringes will tend to vanish as the exiting
wavepackets from the two arms no longer overlap. This yields the
following condition for the exiting wavepackets to not overlap
\begin{equation}\label{con2}
\frac{k_p}{k}>
\sqrt{\frac{N}{2}}\frac{x_{osc}\sqrt{1+\tau^2}}{(x_m-x_0)} ,
\end{equation}
where $x_m$ is the mean position of the recombined wavepackets at
the scaled time $\tau=\omega t$ of detection. Furthermore, if
$\tau\approx \omega (x_m/v_x)\ll 1$, then the atomic wavepacket
will not spread much during its passage through the
interferometer.

\section{Numerical results}\label{results}
In this section we present numerical solutions from our model to
illustrate the interferometric response of the double-X atom
interferometer to a variety of imposed perturbations. These
illustrations are intended to display some interesting features of
the theory and not to serve as detailed study of a particular mode
of inteferometer operation, eg. a rotation sensor.

For our numerical study of the interferometric response of the
double-X interferometer we choose parameters corresponding to the
optical-tweezer loading experiment of Ref. \cite{Gustavson}, with
longitudinal trap frequency $\nu=4 {\rm Hz}\ \Rightarrow
\omega=25.1\ {\rm Hz}$, which gives an oscillator length of $
x_{osc}= 1.04\times 10^{-2}\ {\rm mm}$ for sodium atoms. The
optimal transfer velocity of $70\ {\rm mm/s}$ is chosen to be the
initial velocity $v_x$ of the wavepacket launched into the
interferometer, which corresponds to a wavevector
$k=267x_{osc}^{-1}$. The arm-lengths in the interferometer are
fixed by taking $L_{1}=0.5\ {\rm mm}$, $L_{2}=2.5\ {\rm mm}$ and
the region of detection to be centered at $x_d=3$ mm, $0.5\ {\rm
mm}$ beyond the X-recombiner, see Fig. \ref{Fig1.X-Tonks}. Figure
\ref{Fig2.X-Tonks} shows the density in each of the two
interferometer arms for $N=51$ atoms just before the perturbing
pulse is applied, and the wavepacket is seen to be well contained
between the two X-splitters at $L_1$ and $L_2$. In the following
discussion the output atom densities are calculated around the
point of detection $x_{m}=3$~mm.
\begin{figure}
\includegraphics*[width=\columnwidth,angle=0]{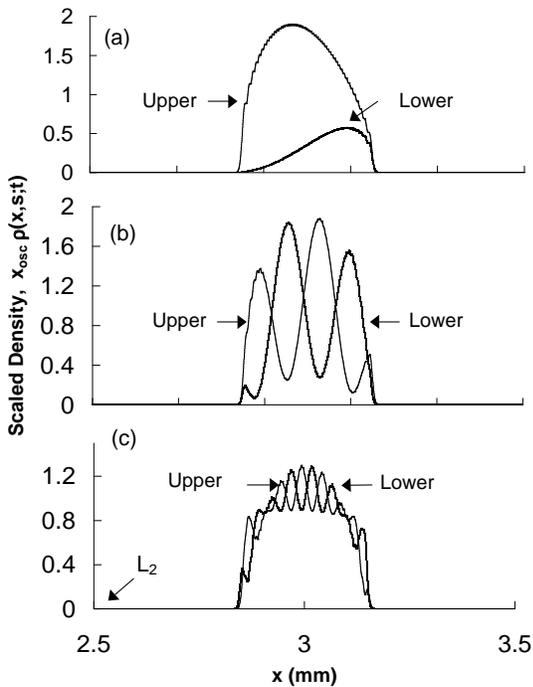}
\caption{Density profile of the recombined wavepackets in the
upper $(s=+1)$ and the lower $(s=-1)$ arms for an initial
wavepacket of $51$ atoms moving with velocity $v_{x}=70$ mm/s, and
the pulsed potentials $V_{s}(x,t)$ applied when $x_{0}=2$ mm: (a)
$k_{p}/k=10^{-4}$ and $Q=0$, (b)$k_{p}/k=10^{-3}$ and $Q=0$ and
(c)$k_{p}/k=10^{-3}$ and $Q/k=5\times 10^{-4}\ \rm{mm}^{-1}$.}
\label{Fig3.X-Tonks}
\end{figure}

The response of the atom interferometer to a net imposed
phase-shift $\varphi$ between the two arms has already been
addressed in the discussion leading to Eq. (\ref{Deln}), so here
we set $\varphi=0$ and look at the effects of an applied force.
Figure \ref{Fig3.X-Tonks} shows the output density profiles in the
upper $(s=+1)$ and the lower $(s=-1)$ arms of the interferometer
for an initial wavepacket of $N=51$ atoms moving with velocity
$v_{x}=70$ mm/s, pulsed force applied at $x_{0}=2$ mm with (a)
$k_{p}/k=10^{-4}$ and $Q=0$, (b)$k_{p}/k=10^{-3}$ and $Q=0$ and
(c)$k_{p}/k=10^{-3}$ and $Q/k=5\times 10^{-4}\ \rm{mm}^{-1}$. For
a spatially uniform force $(Q=0)$ the condition (\ref{con1}) to
observe fringes yields $k_p/k > 0.6\times 10^{-3}$, and this is
consistent with case (a) which shows no fringes $k_p=10^{-4}$, and
case (b) for $k_p=10^{-3}$ which shows a couple of fringes. Thus,
the period of the density fringes in each arm can act as a measure
of the wavevector of the applied force via Eq. (\ref{period}).
However, as shown in Fig. \ref{Fig3.X-Tonks}(c), which is the same
as (b) except with $Q\ne 0$, if the force has a spatially
non-uniform component the visibility fringe can be reduced. We
have found that the density fringes are sensitive to small
spatially inhomogeneous compenents of the applied force. As the
wavevector $k_p$ of the applied perturbation is further increased
beyond that used in Fig. \ref{Fig3.X-Tonks}(b), again with $Q=0$,
the number of density fringes increases but eventually their
contrast decreases. In particular, according to the condition
(\ref{con2}) if $k_{p}/k>0.05$ then the exiting atom wavepackets no
longer overlap and interference fringes are not possible.
\begin{figure}
\includegraphics*[width=\columnwidth,angle=0]{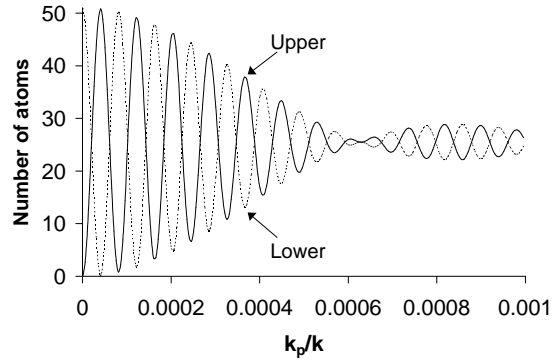}
\caption{Number of atoms in the upper $(s=+1)$ and the lower arms
as a function of linear pulse strength $k_{p}$, with $Q=0$ and
with the time of application of the pulses $V_{s}(x,t)$ fixed at
when the packet centers are at $x_{0}=2$ mm. The initial
wavepacket has $51$ atoms moving with velocity $v_{x}=70$ mm/s.}
\label{Fig4.X-Tonks}
\end{figure}

As a second example, Fig. \ref{Fig4.X-Tonks} shows the response of
the interferometer to a force of varying wavevector $k_p$ for
$x_0=2$ mm and $N=51$ as before. (Keeping the position or time of
application of the applied force fixed is somewhat artificial but
serves to illustrate our theory). In particular, the figure shows
the number of atoms exiting each arm as a function of $k_p$. For
$k_p/k\ll 0.6\times 10^{-3}$, so that there are no density fringes
under the atom wavepacket, the atom numbers exhibit almost
complete periodic oscillations. We can understand this behavior
from Eq. (\ref{lindensity1}) for the densities in each arm: When
$k_p/k\ll 0.6\times 10^{-3}$ the phase $\theta(x,t)$ varies little
over the spatial extent of the atom wavepacket, in which case the
integrated atom numbers in each arm are given to a good
approximation by Eq. (\ref{Npm}) with $\varphi$ replaced by
$2k_px_0$, which shows periodic exchange. In contrast, for $k_p/k>
0.6\times 10^{-3}$ the atom number oscillations are much smaller
in amplitude and centered around 50:50 splitting. In the regime
$k_p/k\gg 0.6\times 10^{-3}$ the phase $\theta(x,t)$ varies
significantly over the spatial extent of the atom wavepacket, and
upon integrating Eq. (\ref{lindensity1}) over $x$ to obtain the
atom numbers in each arm, the cosine terms will tend to integrate
to zero due to their oscillatory nature, giving 50:50 splitting.
\section{Summary and conclusions}
In summary, by introducing a generalization of the Fermi-Bose
mapping for coupled waveguides we have developed the theory of a
double-X atom interferometer in the Tonks-gas limit
\cite{Olshanii,PetShlWal00} of tight transverse confinement and
impenetrable interactions between the atoms. We believe this is a
significant development in view of current efforts to create 1D
integrated atom interferometers, and in waveguide on a chip
technology in general, for inertial and other sensors. In
particular, we have calculated the interferometric response of the
double-X device for a variety of imposed external perturbations to
illustrate the utility of our theory. A key result of our work is
to show that despite the fact that the we are in the Tonks-gas
limit, where the condensed fraction of the initial gas tends to
zero for large number of atoms, the interferometer can still show
high visibility density fringes at the output. That is, our
results show the 1D waveguide devices with ultracold but
non-condensate atom sources can still exhibit a high degree of
first-order coherence. Furthermore, due to the Fermi-Bose mapping
our results apply equally well to a spin-polarized fermionic atom
source, for which s-wave scattering is absent.
\begin{acknowledgments}
We are grateful to Mara Prentiss and Brian Anderson for helpful discussions.
This work was supported by Office of Naval Research
grant N00014-99-1-0806 and the US Army Research Office.
\end{acknowledgments}
\appendix
\section{Time evolution of the Orbitals}
In the absence of external potentials the time evolved orbitals
$u_{n}(x,s;t')$ at time $t=t'+\Delta t$ are obtained from those at
time $t$ by applying the retarded free particle Green's function
\bn u_{n}(x,s;t)=\frac{1}{\sqrt{2\pi i\omega \Delta t}}
\int_{-\infty}^{\infty}dx'\ u_{n}(x',s;t')
e^{-\frac{(x-x')^{2}}{2i\omega \Delta t}} . \en
For brevity in notation notation we introduce the scaled variables
$\tau_{0}=\omega t_{0}$ for the time of application of the pulse,
$\tau=\omega t$ for the time of detection, and $\tau_{d}=\omega
(t-t_{0})$. We also introduce Greek letters to designate the
scaled variables $\chi=x/x_{osc}$ and $\kappa=kx_{osc}$ and
$\Omega=Qx_{osc}^{2}$. The free propagation of each orbital
$u_{n}(x,s;t)$ via either arm is identical up until the instant
$t=t_{0}^{-}$ before the pulses are applied
\bn u_{n}(x,s;t_{0}^{-})=
\frac{C_{n}e^{-i\frac{\kappa^{2}\tau_{0}}{2}}}{
\sqrt{1+i\tau_{0}}}\
\left[\frac{1-i\tau_{0}}{1+i\tau_{0}}\right]^{n/2}\ \h{5mm}\n\\
\times e^{i\kappa \chi}e^{\frac{-(\chi-\chi_{0})^{2}}
{2(1+i\tau_{0})}}H_{n}\left(\frac{\chi-\chi_{0}}{\sqrt{1+\tau_{0}^{2}}}\right)
, \en
where the new center of the wavepacket is at
$\chi_{0}=\kappa\tau_{0}$.  After the pulsed potentials
$V_{s}(x,t)$ are applied the orbitals acquire an additional
space-dependent phase different in each arm
\begin{equation}
u_{n}(x,s;t_{0}^{+})=e^{is(\kappa_{p}\chi+\Omega\chi^{2})}u_{n}(x,s;t_{0}^{-}).
\end{equation}
Evaluating the subsequent free propagation till a time $t>t_{0}$
{\it after} the wavepacket from each arm has completely passed
through the recombiner, one finds the functional form of the modes
coming from the two arms to be
\bn \label{quadratic}u_{n}(x,\pm 1;t)= \frac{C_{n}
e^{-\frac{i}{2}\kappa^{2}\tau_{0}}} {
\sqrt{(1\pm2\Omega\tau_{d})+i(\tau\pm 2\tau_{0}\Omega\tau_{d})}}
\h{5mm}\n\\ \times\left[\frac{(1\pm
2\Omega\tau_{d})-i(\tau\pm2\tau_{0}\Omega\tau_{d})}
{(1\pm2\Omega\tau_{d})+i(\tau\pm2\tau_{0}\Omega\tau_{d})}\right]^{n/2}
e^{\frac{F_{\pm}(\chi,\tau,\tau_{0})}{G_{\pm}(\chi,\tau,\tau_{0})}}\
\n\\ \times
H_{n}\left(\frac{\chi-[\chi_{0}+(\kappa_{\pm}\pm2\chi_{0}\Omega)\tau_{d}]}
{\sqrt{[1\pm2\Omega\tau_{d}]^{2}+[\tau\pm2\tau_{0}\Omega\tau_{d}]^{2}}}\right)
, \h{2mm}\en
with $\kappa_\pm=\kappa\pm \kappa_p$, and the argument of the
exponential is given by
\bn F_{\pm}(\chi,\tau,\tau_{0})=
-(1\pm2\Omega\tau_{d})\chi_{0}^{2}-\kappa_{\pm}\tau_{d}[2\chi_{0}+i\kappa_{\pm}(1+i\tau_{0})
]\n\\+2\chi[
\chi_{0}+i\kappa_{\pm}(1+i\tau_{0})]-[(1\pm2\Omega\tau_{0})-i(\pm2\Omega)]\chi^{2}
\n\\
G_{\pm}(\chi,\tau,\tau_{0})=2[(1+i\tau)+(\pm2\Omega)\tau_{d}(1+i\tau_{0})]
. \h{1.8cm}\n\en
When the potential has only a linear dependence on $x$ i.e.
$\Omega=0$ this expression acquires a more transparent form
\bn \label{linearu} u_{n}(x,\pm 1;t)=
\frac{C_{n}e^{-\frac{i}{2}(\kappa^{2}_{\pm}\tau_{d}
+\kappa^{2}\tau_{0})} } { \sqrt{1+i\tau}} \left[\frac{1-i\tau}
{1+i\tau}\right]^{n/2}\ \n\\ \times
e^{i\chi\kappa_{\pm}}e^{\frac{-[\chi-[\kappa_{\pm}\tau_{d}+
\chi_{0}]]^{2}}{2(1+i\tau)}}H_{n}\left(\frac{\chi-(\chi_{0}+\kappa_{\pm}\tau_{d})}
{\sqrt{1+\tau^{2}}}\right).\h{2mm} \en
This expression contains a phase-independent factor which is a
Hermite-Gaussian centered at $x_{\pm}=x_{0}+\kappa_{\pm}\tau_{d}$
with width $\sqrt{1+\tau^{2}}$, that we represent in terms of
unscaled variables
\bn \label{linearu1} U_{n}(x\!-\!x_{\pm},t)\!\!=\!\!\frac{C_{n}\
e^{\frac{-[x-x_{\pm}]^{2}}{2[1+(\omega^{2} t)^{2}]}} } {\
[1\!\!+\!\!(\omega t)^{2}]^{1/4}}
H_{n}\!\left(\!\!\frac{x-x_{\pm}} {x_{osc}\sqrt{1\!\!+\!(\omega
t)^{2}}}\!\!\right).\h{2mm} \en
\end{document}